\documentclass[10pt]{iopart}
\usepackage{psfig}
\begin{document}

\title[Subthreshold kaon production and nuclear medium]{
Kaon production at subthreshold energies\\
what do we learn about the nuclear medium ?  
\footnote[1]{Supported by IN2P3/CNRS and GSI.}}

\author{C Hartnack\footnote[2]{invited speaker}$^1$, H Oeschler$^2$   
and J Aichelin$^1$ }

\address{
$^1$SUBATECH,  UMR 6457, Ecole des Mines de Nantes, IN2P3/CNRS et Universit\'e de
Nantes,  4, rue A. Kastler, B.P. 20722,
44307 Nantes, France \\
$^2$ Institut f\"ur kernphysik, Darmstadt University of Technologie, 
64289 Darmstadt, Germany}

\begin{abstract}
The IQMD model is used to compare spectra and elliptic flow of 
kaons produced at subthreshold energies with data taken at the SIS
accelerator at GSI.
We find that temperatures of the spectra are dominated by the rescattering
of the kaons. The study of elliptic flow observables indicates the influence
of rescattering as well as of the optical potential of the kaons with increasing
dominance of the optical potential at lower incident energies.

\end{abstract}

\vspace*{-3mm}

\section{Introduction}
One key question in the analysis of sub-threshold kaon production 
is how to obtain information on the properties of strange mesons in dense nuclear 
matter. Especially the  relation of the optical potential of $K^+$ 
and $K^-$ in the nuclear medium  to experimental observables like the
$K^-/K^+$ ratios, mesonic in-plane flow and azimuthal distribution of kaons
is subject of vivid discussions, who have triggered a lot of activities on the
experimental 
\cite{Barth,Menzel,Ritman,Crochet,Devismes,Laue,Sturm,Foerster,uhlig,FoersterPRC}
and theoretical side 
\cite{schaffi,Cassing,CLE00,ho_s2000,Ko01,Fuchs,LiKo,Wang,oldhartnack}.

In this article we study the production of $K^+$ in the reaction Au+Au at 
1.48 AGeV which corresponds to recent experiments performed by 
the KaoS collaboration \cite{Foerster,FoersterPRC}.
 For this purpose we use the IQMD model \cite{iqmd,bass}
where we have supplemented our standard simulation program by  all 
relevant cross sections for kaon production and annihilation and a 
(density and momentum dependent) $KN$ optical potential.  
For the latter we use a parametrization resulting from 
relativistic mean field calculations of Schaffner-Bielich \cite{juergen}. 
Detailed description of our simulations as well as the discussion of
the effects of the productions cross sections and the optical potentials
on the yields of strangeness production can be found in 
\cite{sqm2001,kminus,sqm2003}.
A more detailed analysis of the presented studies as well as analysis of 
other systems and observables will be published \cite{bigpr}.

\section{Kaon spectra}
\begin{figure}[hbt]
 \centerline{\psfig{file=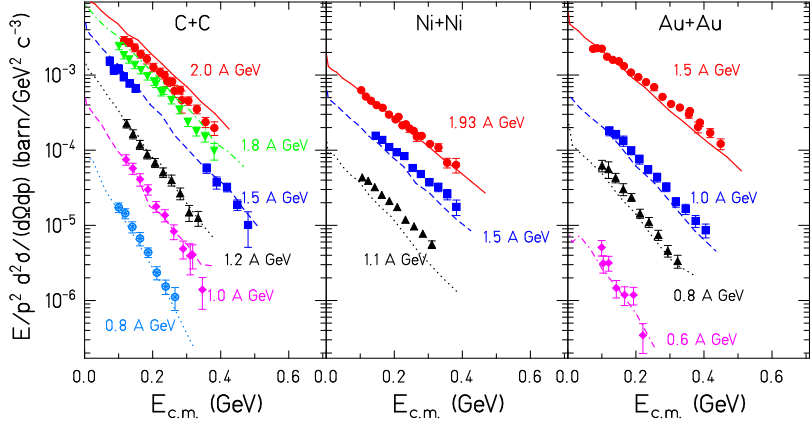,width=0.9\textwidth,angle=-0} }
\caption{Comparison of $K^+$ spectra measured by the KaoS collaboration 
in Au+Au, Ni+Ni and C+C collisions at different energies  
with IQMD calculations.}
\vspace*{-0.3\baselineskip}
\label{kaoscompa}
\end{figure}

In the past decades FOPI and KaoS experiments at GSI measured the production of strange
particles at various  combinations of system size and incident energy.
Figure \ref{kaoscompa} presents a systematics of the invariant spectra of kaons 
performed by the KaoS collaboration in C+C, Ni+Ni and C+C collisions 
\cite{FoersterPRC} (shown as symbols) with IQMD 
calculations using our standard rescattering and optical potential
(lines).\vspace*{-0.2\baselineskip}

We see that the IQMD model is well able to describe the experimental data
over the full range of energy and system size. 
We note that in the energy range of the data (about 100-500 MeV
in the centre of masss frame) the spectra are quite comparable to thermal 
spectra with one  temperature, rising up with energy and system size.
  
Our aim is to study the influence of medium effects on the spectra of kaons.
These effects may be caused by hard collisions with the nuclear medium
(rescattering) or by a soft interaction (optical potential).
In the following analysis we concentrate on the heaviest system Au+Au,
in the expectation to find here the strongest in-medium effects. 
Furthermore we focus on the highest energy (1.48 AGeV) which gives us the
possibility to yield high statistic $K^+$ as well as $K^-$.

\begin{figure}[hbt]
 \centerline{\psfig{file=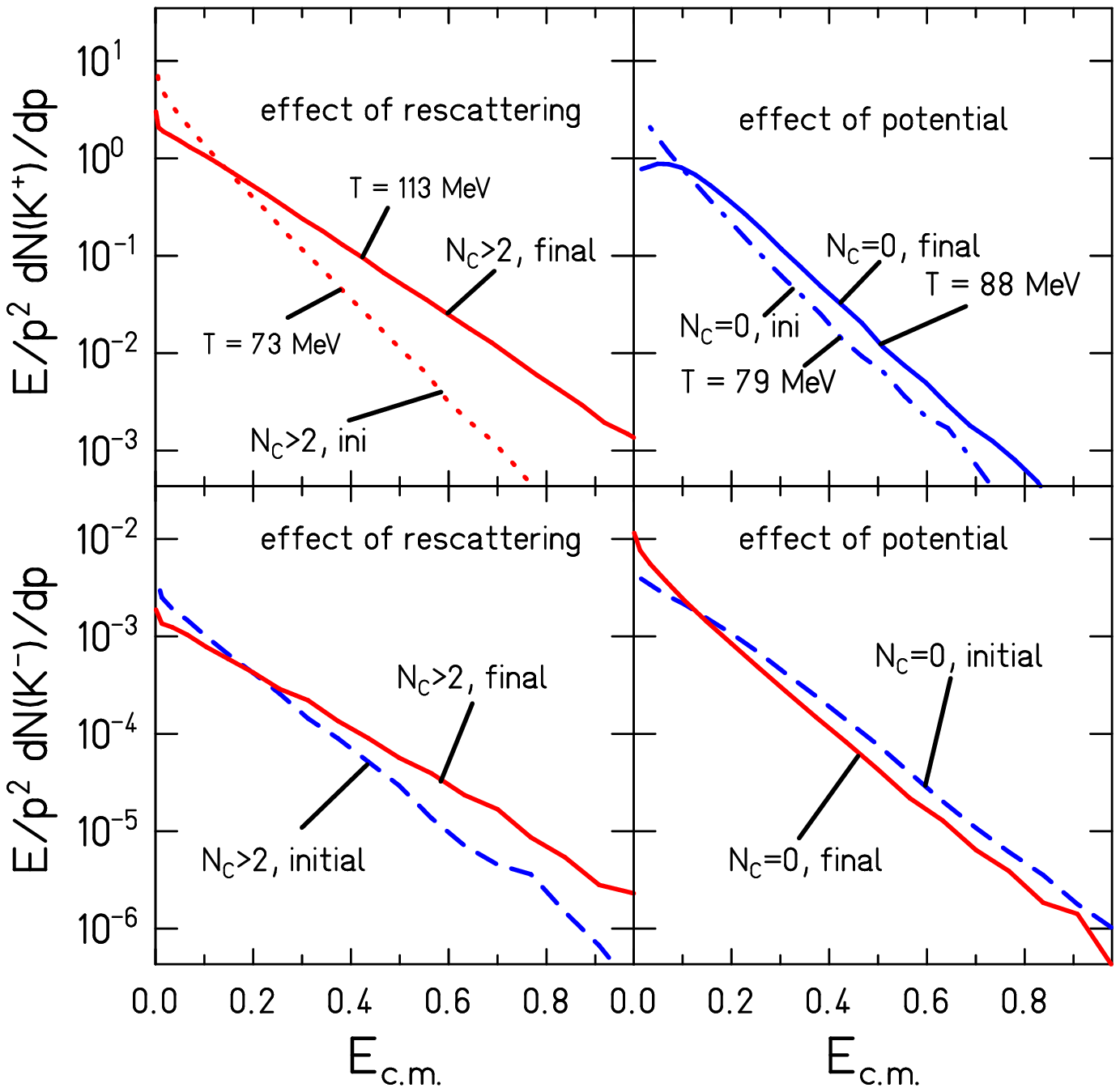,width=0.65\textwidth,angle=-0}
\hspace*{-0.06\textwidth}
\psfig{file=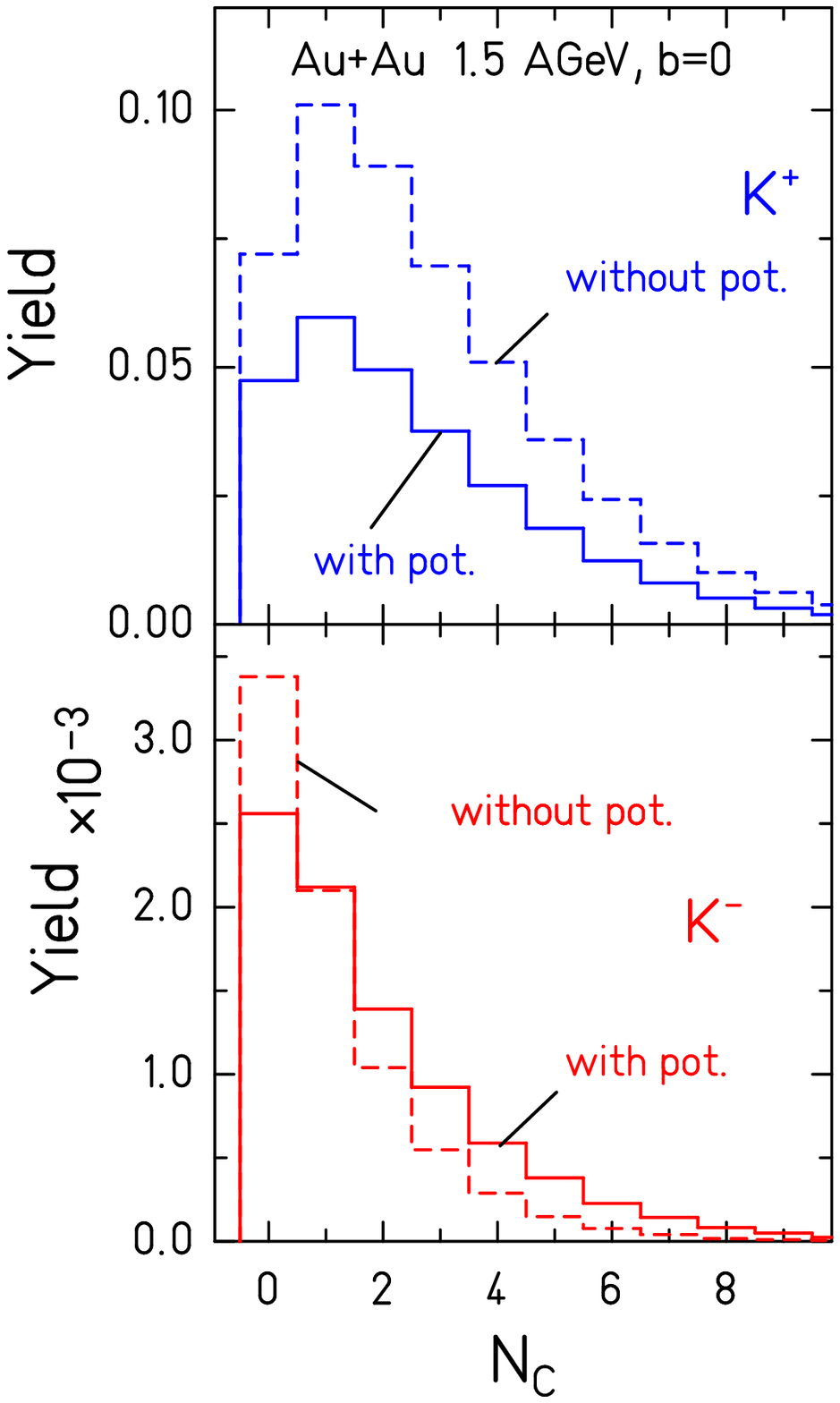,width=0.4\textwidth,angle=-0} }
\vspace*{-0.3\baselineskip}
\caption{Left: Invariant spectra of $K^+$ (above) and $K^-$ (below) for 
the final state (full line) and directly after production (dashed/dotted lines)
for particles that felt many collisions but no potential (left) and 
those who felt an optical potential but no collisions (middle).
Right: Distribution of the number of rescatterings for $K^+$ (above) 
and $K^-$ (below) in calculations with (full line) and without (dashed) 
an optical potential for the kaons.
 }
\label{spectra-coll}
\end{figure}

Figure \ref{spectra-coll} presents a detailed analysis of the contribution 
of rescattering and optical potential to the spectra.
The upper row describes the spectra of $K^+$ while the lower row describes
the same analysis for the $K^-$.
The left part of the figure shows an analysis for particles which scattered
quite frequently ($N_C>2$) stemming from calculations where no optical potential
was implemented. The effects are thus purely related to rescattering.
The dotted (dashed) line shows the analysis of the momenta directly after
production (i.e. before the rescattering happened) while the full line 
describes the final momenta (after rescattering). We see that as well for
$K^+$ as for $K^-$ the rescattering with the surrounding nuclear matter
heats up the kaons yielding a significantly higher temperature.

The middle part of figure \ref{spectra-coll} shows the effect of the optical
potential, which is repulsive for  $K^+$ and attractive for $K^-$. 
Again we compare the momenta
of the kaons directly after their production (dashed/dashed-dotted line) with
the final momenta (full line). For this analysis we focus on particles which did not scatter. Thus the signal
can be interpreted as a potential effect.
We see that the potential changes the spectra only slightly. Due to the
repulsion of $K^+$ their temperatures get slightly enhanced (upper row) while the 
temperatures of the $K^-$ (lower row) are slightly reduced due to the attraction 
of the potential.
However the effect of the potential is much weaker than the effect of the rescattering
shown on the left side, both for $K^+$ (upper part) and $K^-$ (lower part).

In order to understand the final kaon spectra quantitatively we have to know
the rescattering quantitatively.
This information is given on the right side of  figure \ref{spectra-coll}.
We see that a large fraction of the $K^+$ (upper row) underwent a significant
number of collisions. Therefore, the spectra are dominated by 
rescattering.  For the $K^-$ (lower row) this contribution is weaker but still
quite noticable.  
 
\begin{figure}[hbt]
 \centerline{
\psfig{file=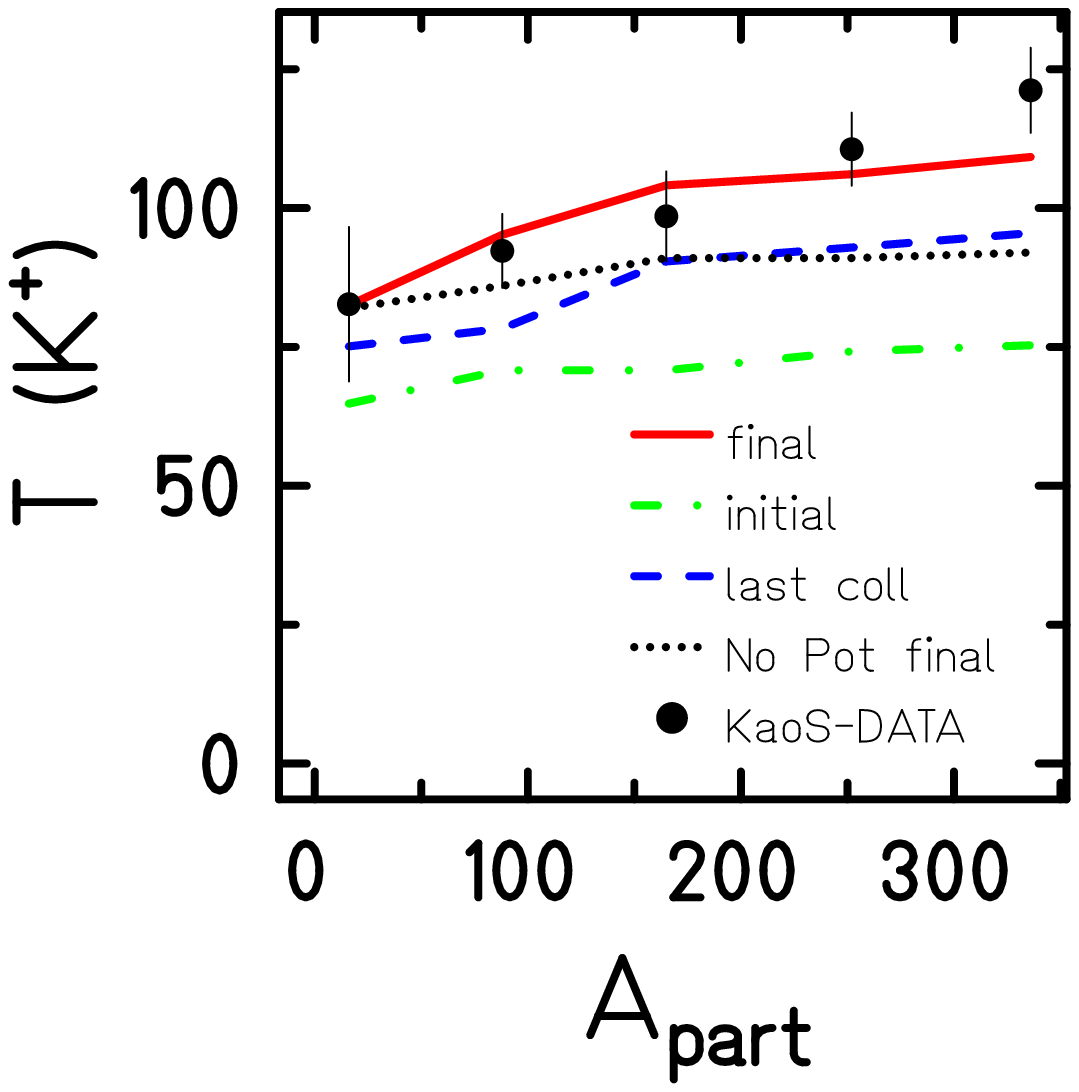,width=0.45\textwidth,angle=-0}
 \hspace*{-0.06\textwidth}
 \psfig{file=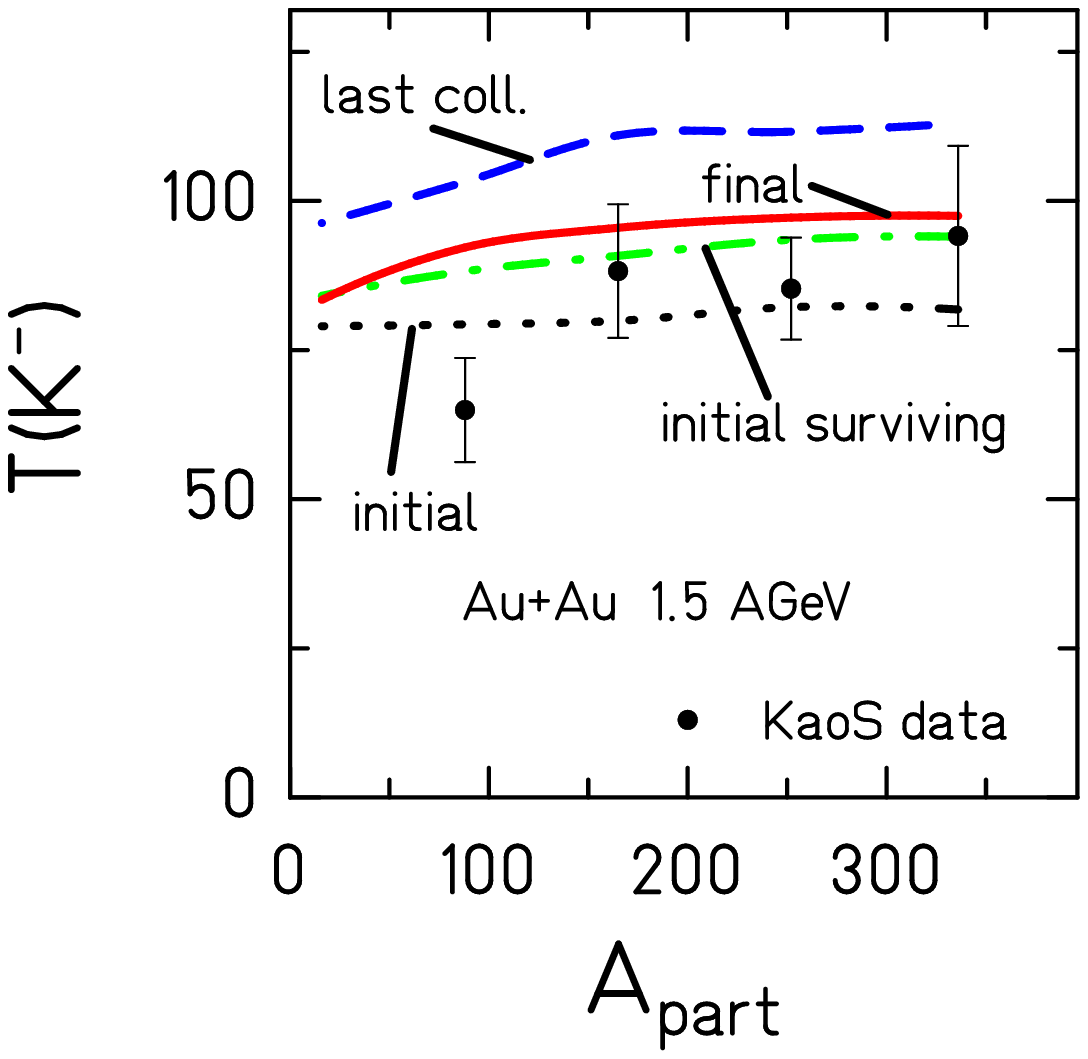,width=0.6\textwidth,angle=-0}}
\vspace*{-0.3\baselineskip}
\caption{Comparison of the centrality dependence of the temperature of 
$K^+$ (left) and $K^-$ (right) between KaoS data (bullets) and
IQMD calculations (lines). For the calculations the kaons have been
analysed directly after their production (dash-dotted line), after
their last collision (dashed line) and in the final state (full line).}
\label{apart}
\end{figure}

Even if the values of the temperatures are majorly due to the
rescattering in the medium, the optical potential still adds a supplementary
push for the $K^+$ and an supplementary attraction for the $K^-$.
This can be seen in figure \ref{apart} where we compare the centrality
dependence of the temperatures in Au+Au collisions at 1.48 AGeV
for $K^+$ (left) and $K^-$ (right) between 
KaoS data (bullets) and IQMD calculations (lines).

Let us start with the $K^+$ (left side): directly after their production
their momenta show a quite low temperature (dash-dotted line). This 
temperature is increased by rescattering. However, still directly after
their last collision (dashed line) the temperature is not reaching its
final value (full line). This is due to the extra-push given by the potential.
A calculation without potential (dotted line) yields about the same
temperatures than kaons  analysed directly after their last collision
in a calculation with optical potential (dashed line).
This can be interpreted that rescattering is "wiping out" the information on the
previous properties of the kaon. Only after the last contact, the optical
potential may leave a signature on the spectra. The significance of this
trace depends on the time when the last contact has happened. 

For the $K^-$ we have to take care of another aspect: the absorption.
Since its cross section is huge at small relative momenta the absorption
"eats up" the low energy kaons and thus imitates higher temperatures.
This can be seen on the right side of figure \ref{apart}:
if we take all produced kaons directly after their production (dotted line)
and compare their temperature to the temperature of the surviving kaons
after their production (dash-dotted line) we see a significant enhancement.
This temperature is still enhanced by (elastic) rescattering. The temperature
of the $K^-$ analysed directly after their last collision (dashed line, rhs)
even depasses the respective temperature of the $K^+$ after their
last collision (dashed line, lhs). However the attractive potential 
reduces the final temperature (full line, rhs) to values below the
respective temperature of the $K^+$ (full line, lhs).
The effect of the final push of the potential is more significant for the $K^-$
than for the $K^+$. This may be due to a stronger density dependence of
optical potential of the $K^-$. However, also the different contribution
of collision numbers (less collisions for $K^-$) might play a role.
We conclude that even if the final temperatures are dominated by rescattering
some traces of the potentials may be left.

\section{Elliptic flow and $v_2$}
Another observable expected to trace information on the optical potential of kaons
is the analysis of elliptic flow commonly described by the Fourier coefficents
$v_1$ (directed flow in the reaction plane) and $v_2$. A negative value of
$v_2$ indicates a flow preferentially out of plane while for a positive
$v_2$ the particles are dominantly emitted inside the reaction plane.
For the reaction Au+Au at 1.48 AGeV incident energy the KaoS collaboration 
measured a out-of-plane emission of kaons at midrapidity \cite{Foerster}.
The azimuthal distribution of the kaons could be well described by the
IQMD model\cite{sqm2003}. However, it was found that the signal is sensitive to the
optical potential of the kaons as well as to their rescattering.
In this article we want to demonstrate that these effects are also related to
the geometry and time evolution of the system.\vspace*{-0.2\baselineskip}

\begin{figure}[hbt]
\centerline{\psfig{file=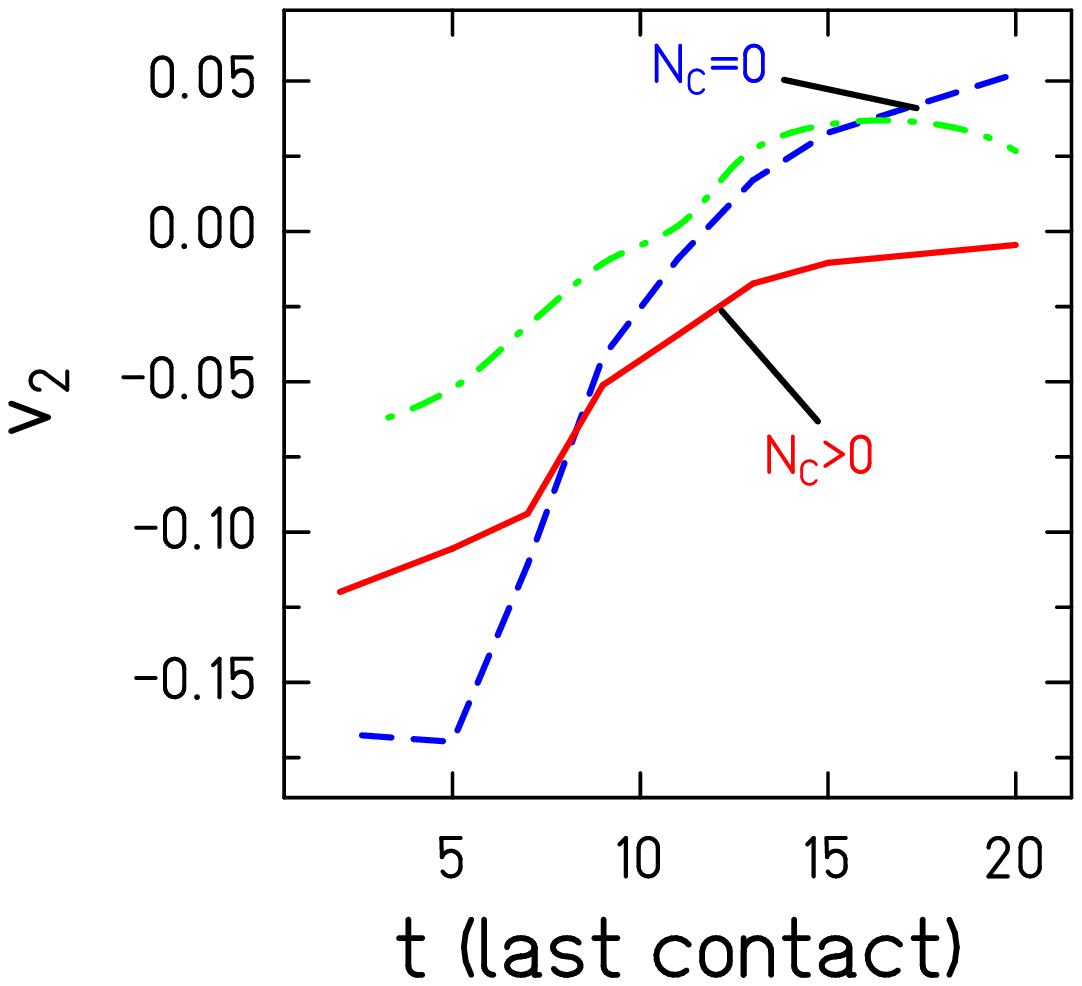,width=0.5\textwidth,angle=-0}
\hspace*{-0.05\textwidth}
\psfig{file=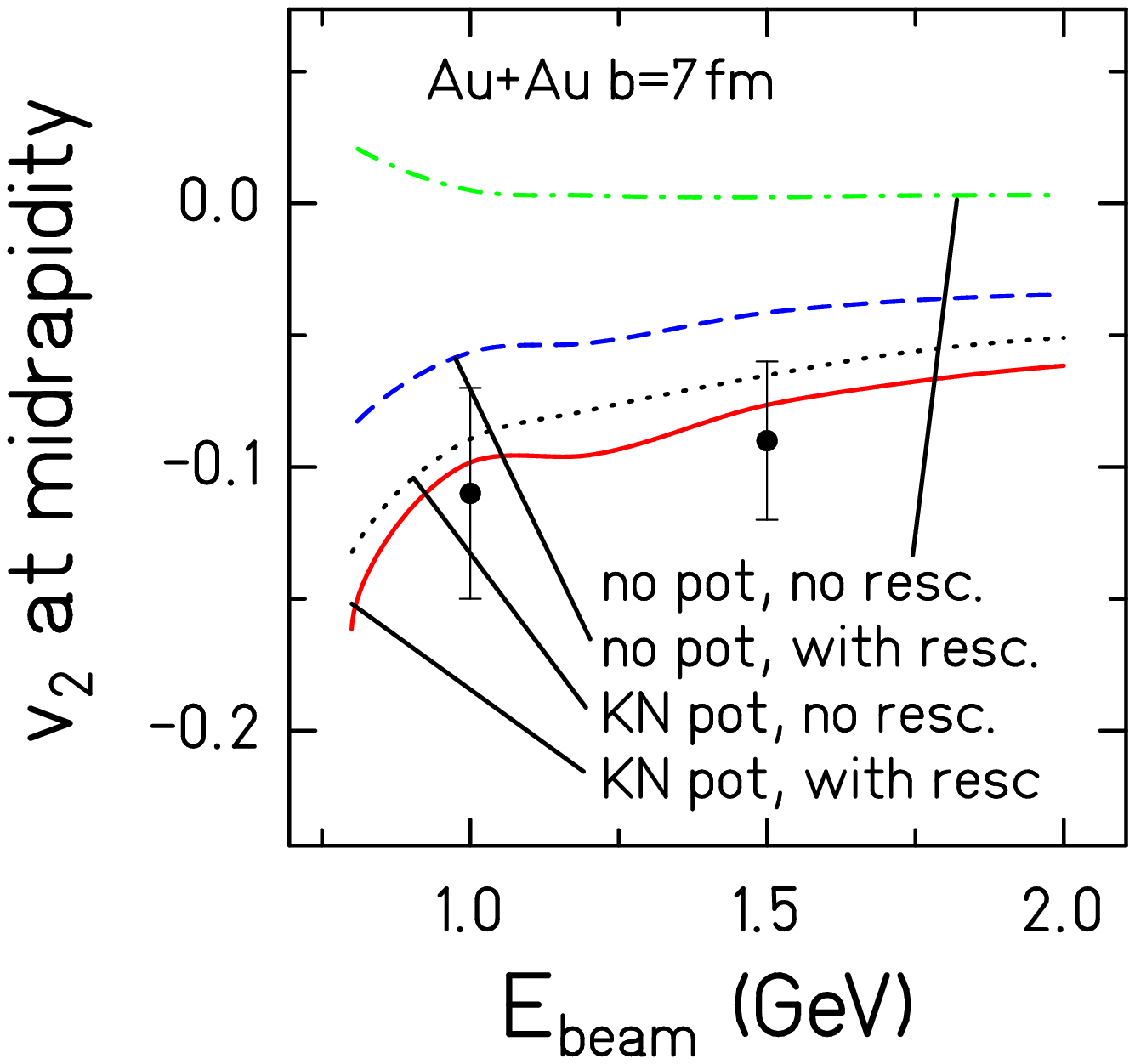,width=0.5\textwidth,angle=-0} }
\vspace*{-0.3\baselineskip}
\caption{Left: The influence of the time of last contact on $v_2$ for the
momenta directly after the contact (dash-dotted line) and for the final
momenta (dashed line) of particles which did not collide and of the final
momenta of particles that collided (full line).
Right: comparison of $v_2$ in a semiperipheral collision of Au+Au at different incident
energies between KaoS data and calculations with different options on optical
potential and rescattering.   }
\label{v2au15}
\end{figure}

The left side of figure \ref{v2au15} analyses the $K^+$ according to the time
when the kaon had his last hard contact (production or rescattering) with the
nuclear matter. For particles that did not collide we see a change of the
sign of $v_2$ for their momenta directly after production (dash-dotted line)
showing a squeeze-out at early times. 
This effect may be due to the geometry of the system, which changes in time.
In the early stage,  due to the presence of the spectator matter, there is a 
higher chance to escape out-of-plane without having a collision. This
bias changes at later times when the spectators moved off.
Afterwards the optical potential pushed the kaons further into the 
chosen direction . 
The final momenta (dashed line) demonstrate that this push is stronger at 
early times, when the densities are high, while for later
times the potential push becomes weaker.  The final momenta of particles that
collided (full line) shows a less significant $v_2$ at early times and approaches
to zero at late times. We find that the potential becomes quite significant for
particles which leave the system quite early.

The right side of figure \ref{v2au15} shows the excitation function of $v_2$ 
at midrapidity in semiperipheral collision of Au+Au. The lines show calculations of 
IQMD with and without rescattering and/or optical potential.
A calculation without potential nor rescattering (dash-dotted line) yields 
$v_2=0$. A calculation with rescattering but without optical potential (dashed line)
shows less significant effect  than a calculation with potential but without
rescattering (dotted line). The best agreement with data (bullets) can, however,
be obtained when activating potentials and rescattering (full line).
The effect of the potential becomes more evident at lower incident energies.   
This is caused by the interplay of emission time and passing time of the spectator matter.

\section{Conclusion}
We analysed the spectra and temperatures of $K^+$ and $K^-$ and the 
elliptic flow of $K^+$ in Au+Au collisions at 1.48 AGeV incident energy.
We find the temperatures being dominated by rescattering with the nuclear
matter. However, the optical
potential is giving an final push of opposite sign for $K^+$ and $K^-$ 
yielding different temperatures of the two species.
The elliptic flow is effected by rescattering and optical potential.
The sensitivity to the optical potential is especially enhanced
for kaons leaving the system quite early.
In reactions at low incident energy the potential plays an increasing role.

\section*{Acknowledgements}
The authors acknowledge a fruitful collaboration with the KaoS experiment,
especially with  A. Foerster and F. Uhlig.
This work was in part supported by the GSI-IN2P3 agreement.


\section*{References}

\begin{thebibliography}{99}
\bibitem{Barth} R. Barth et al., (KaoS  Collaboration),
Phys. Rev. Lett. {\bf 78} (1997) 4007.
\bibitem{Menzel} M.~Menzel et al., (KaoS  Collaboration),
Phys. Lett. {\bf B 495} (2000) 26;\\
M. Menzel, Dissertation, Universit\"at Marburg, 2000.
%\bibitem{Ahle}L.~Ahle et al., (E802 Collaboration), Phys. Rev. {\bf C58} (1998) 3523.
\bibitem{Ritman} J.L. Ritman et al. (FOPI collaboration),  Z. Phys. {\bf A352} (1995) 355
\bibitem{Crochet}  P. Crochet et al. (FOPI collaboration), Phys. Lett. {\bf B486} 
(2000) 6
\bibitem{Devismes} A. Devismes et al. (FOPI collaboration),
J. Phys. G {\bf 28} (2002) 1591.
\bibitem{Laue} F. Laue %, C. Sturm
et al., (KaoS  Collaboration), Phys. Rev. Lett. {\bf 82} (1999) 1640.
\bibitem{Sturm} C. Sturm et al (KaoS Collaboration), Phys. Rev. Lett. {\bf 86}
(2001) 39 \\
C. Sturm et al (KaoS Collaboration), J. Phys. G {\bf 28} (2002) 1895
\bibitem{Foerster} A.~F\"orster (KaoS  Collaboration), PhD thesis, TH Darmstadt;
A. F\"orster et al. Phys. Rev. Lett. {\bf 31} (2003) 152301;
A. F\"orster et al. J. Phys. G. {\bf 30} (2004) 393
 %, C. Sturm
\bibitem{uhlig} F. Uhlig et al., (KaoS collaboration),  Phys.Rev. Lett. {\bf 95}
(2005) 012301
\bibitem{FoersterPRC} A.~F\"orster (KaoS  Collaboration), Phys. Rev. {\bf C 75} (2007) 024906
\bibitem{schaffi} J. Schaffner-Bielich et al., Nucl Phys {\bf A669} (2000) 153; 
\bibitem{Cassing}W.~Cassing et al., Nucl.~Phys.~A{\bf 614} (1997) 415. \\
E. Bratkovskaya et al., Nucl. Phys. {\bf A622} (1997) 593 \\
W. Cassing and E. Bratkovskaya, {\it Phys. Rep} {\bf 308} (1999) 65 
\bibitem{CLE00} J. Cleymans, H. Oeschler and K. Redlich,
Phys. Lett.~{\bf B485} (2000) 27.
\bibitem{ho_s2000} H.~Oeschler, J.~Phys.~G:  {\bf 27} (2001) 257.
\bibitem{Ko01} 
C.M. Ko, G.Q Li, J. Phys. G {\bf 22} (1996) 1673. \\
C.M. Ko, J.~Phys.~G {\bf 27} (2001) 327.
\bibitem{Fuchs} 
C. Fuchs et al. Phys. Lett {\bf B 434} (1998) 245, \\
C. Fuchs et al. Phys. Rev. Lett {\bf 86} (2001) 1794 \\
C. Fuchs et al. J. Phys. G {\bf 28} (2002) 1615 
\bibitem{LiKo} 
G Q Li et al.  Phys. Lett {\bf B 381} (1996) 17 
\bibitem{Wang} 
Z.S. Wang et al. Eur. Phys. J {\bf A 5} (1999) 275 
\bibitem{oldhartnack} C. Hartnack et al., Nucl. Phys. {\bf A580} (1994) 643.
\bibitem{iqmd}C. Hartnack et al., Eur.Phys.J. {\bf A1} (1998) 151.
\bibitem{bass} S.A. Bass et al. Phys. Rev. {\bf C50} (1994) 2167.\\ 
S.A. Bass et al. Phys. Rev. {\bf C51} (1995) 3343.
\bibitem{juergen}J. Schaffner et al., Nucl. Phys. {\bf A625} (1997) 325.
\bibitem{sqm2001}
C. Hartnack and J. Aichelin, J. Phys. G {\bf 28} (2002) 1649
\bibitem{kminus}
C. Hartnack, H. Oeschler and J. Aichelin,  Phys.Rev.Lett. {\bf 90} (2003) 102302.  
\bibitem{sqm2003} C. Hartnack and J. Aichelin, J. Phys. G {\bf 30} (2004) 531
\bibitem{bigpr} C. Hartnack, H. Oeschler and J. Aichelin, Phys. Rep. in preparation
\end{thebibliography}
\end{document}